\documentclass[aps,showpacs,showkeys,superscriptaddress]{revtex4}

\usepackage{graphicx}

\begin{document}
\title{Green Function Approach for Calculation of the Local Density of States in the
Graphitic Nanocone}

%\author{Jan Smotlacha\inst{1,2}\fnsep\thanks{\email{smota@centrum.cz}} \and
%        Richard Pin\v{c}\'{a}k\inst{1,3}\fnsep\thanks{\email{pincak@saske.sk}}
%      }

%\institute{Bogoliubov Laboratory of Theoretical Physics, Joint
%Institute for Nuclear Research, 141980 Dubna, Moscow region, Russia
%\and
%           Faculty of Nuclear Sciences and Physical Engineering, Czech Technical University, Brehova 7, 110 00 Prague,
%Czech Republic
%\and
%           Institute of Experimental Physics, Slovak Academy of Sciences, Watsonova 47, 043 53 Kosice, Slovak Republic}

\author{J. Smotlacha}\email{smota@centrum.cz}
\affiliation{Faculty of Nuclear Sciences and Physical Engineering, Czech Technical University, Brehova 7, 110 00 Prague,
Czech Republic}

\author{R. Pincak}\email{pincak@saske.sk}
\affiliation{Bogoliubov Laboratory of Theoretical Physics, Joint
Institute for Nuclear Research, 141980 Dubna, Moscow region, Russia}
\affiliation{Institute of Experimental Physics, Slovak Academy of Sciences,
Watsonova 47,043 53 Kosice, Slovak Republic}

\begin{abstract}
Graphene and other nanostructures belong to the center of interest of today's physics research. The local density of states of the graphitic nanocone influenced by the spin--orbit interaction was calculated. Numerical calculations and the Green function approach were used to solve this problem. It was proven in the second case that the second order approximation is not sufficient for this purpose.
\end{abstract}

\pacs{73.22.Pr, 81.05.ue, 71.70.Ej, 72.25.-b.}

\keywords{Tight-binding method,
Graphitic nanocone, Spin--orbit coupling, Green function approach}

\maketitle
\section{Introduction}
\begin{figure}[h]
\centering
\includegraphics[width=43mm]{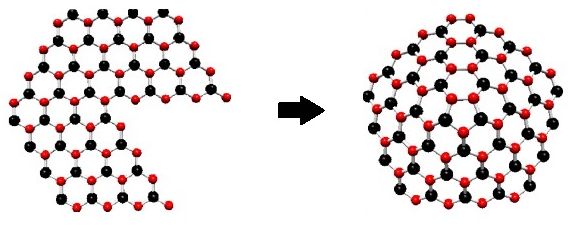}
\hspace{15mm}
\includegraphics[width=20mm]{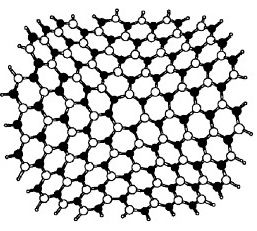}
\caption{Creation of the defects in the graphene structure: pentagons for positive curvature (left), heptagons for the negative curvature (right)}\label{defects}
\end{figure}

\
\label{intro}
The chemical structure of the carbon nanomaterials is based on the hexagonal atomic carbon lattice. The basic structure of this kind is graphene, a plain carbon monolayer. The molecules of other forms of this kind have this structure disrupted mostly by the pentagonal or the heptagonal defects. They can be created from the graphene monolayer by cutting or adding $60^{\circ}$ sectors (Fig.~\ref{defects}).

\begin{figure}[h]
\centering
\includegraphics[width=30mm]{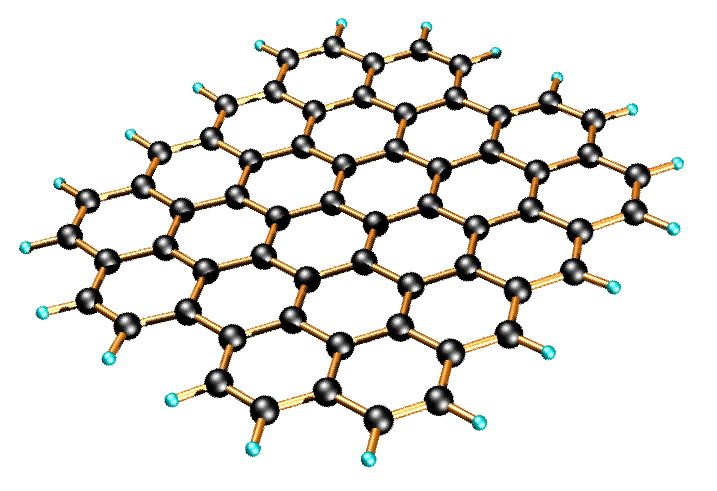}
\hspace{10mm}
\includegraphics[width=23mm]{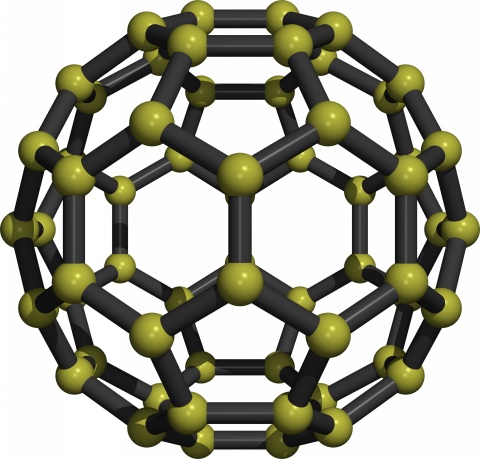}
\hspace{10mm}
\includegraphics[width=27mm]{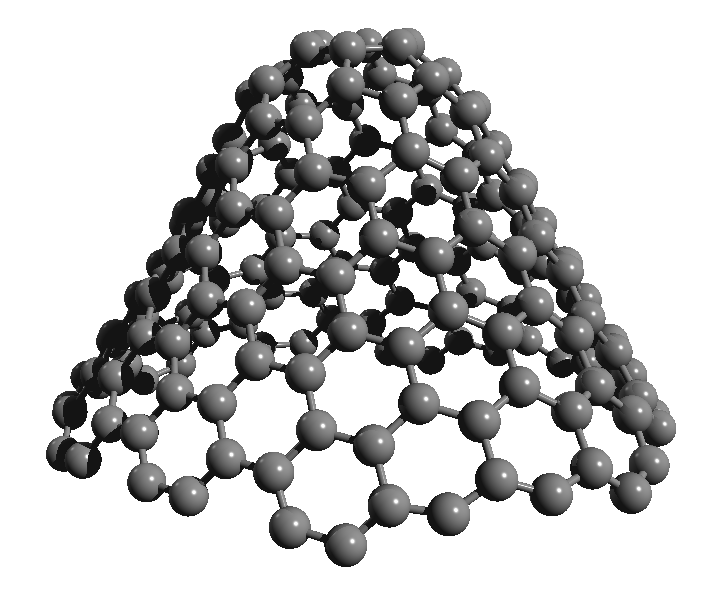}
\caption{Different kinds of the nanostructures: graphene (left), fullerene (middle), nanocone (right)}\label{nano}
\end{figure}

The name of the derived forms is mostly given by the shape -- we have (see Fig.~\ref{nano}) nanotubes, nanocones, nanowires, nanoribbons, nanotoroids, etc.  The well-known form is the fullerene which can be considered as a nanosphere. The electronic properties of all these materials make them interesting for physicists -- they are good candidates for the construction of the molecular electronic devices to computers.

\section{Electronic structure}

The electronic structure is characterized by the local density of states (LDoS) -- the number of electronic states per the unit surface area and the unit interval of energies. For the purpose of its calculation one of these two methods is mostly used -- the direct calculation from the electronic spectrum \cite{wallace, slonczewski} (the case of the periodical structures like graphene, fullerene, etc.), or, using the continuum gauge field-theory approach, a Dirac-like equation is solved \cite{mele} (the aperiodical structures like wormhole, nanocone, etc.).

In this paper, we use the second method. The corresponding Dirac--like equation for the massless fermion \cite{kochetov, smotlacha} has the form
\begin{equation}\label{DirEq}{\rm i}\sigma^{\alpha}e_{\alpha}^{\mu}[\partial_{\mu}+\Omega_{\mu}-{\rm i}a_{\mu}-{\rm i}a_{\mu}^W-{\rm i}A_{\mu}]\Psi^T=E\Psi^T,\hspace{5mm}
 \Psi^T=\left(\begin{array}{c}F_1 \\ F_2\end{array}\right).\end{equation}
Here, the particular terms have this meaning: $\sigma^{\alpha}$ are the Pauli matrices and $e_{\alpha}^{\mu}$ is the zweibein \cite{fecko} which can be defined with the help of the metric tensor
\begin{equation}
g_{\mu\nu}(x)=e^{\alpha}_{\mu}(x)e^{\beta}_{\nu}(x)\eta_{\alpha\beta}\,,
\end{equation}
where $\eta_{\alpha\beta}$ is the metric of the uncurved space. Next, $\omega_{\mu}$ is the spin connection (a kind of a connection 1-form, \cite{fecko}) and for the case of the rotational symmetry we have
\begin{equation}\omega^{12}_{\varphi}=-\omega^{21}_{\varphi}=1-\frac{\partial_{r}\sqrt{g_{\varphi\varphi}}}
{\sqrt{g_{rr}}}=2\omega,\hspace{1cm}\omega^{12}_{r}=\omega^{21}_{r}=0.\end{equation}
Then, $\Omega_{\mu}=\frac{1}{8}\omega^{\alpha\beta}_{\mu}[\sigma_{\alpha},\sigma_{\beta}]$ is the spin connection in the spinor representation. The gauge fields $a_{\mu}, a_{\mu}^W$ ensure the circular periodicity:
\begin{equation}a_{\varphi}=N/4,\hspace{1cm}a_{\varphi}^W=-\frac{1}{3}(2m+n),\end{equation}
where $N$ is the number of the defects and $(n,m)$ is the chiral vector. The last term $A_{\mu}$ represents the electromagnetic potential coming from a possible presence of the magnetic field.

Then, for the given energy, the LDoS is got as the sum of the squares of the radial components of the wave function,
\begin{equation}\mathrm{LDoS}=F_1^2+F_2^2.\end{equation}

\section{Graphitic nanocone}

In the case of the graphitic nanocone, which is the aperiodical structure, we use the formalism of the last section for calculation of its electronic structure. For the description of the geometry, we use the coordinates $r$ (radial) and $\varphi$ (angular) and denote $\eta=N/6$. Then the radius-vector is
\begin{equation}
\vec{R}=\Big(r(1-\eta)\cos\varphi,r(1-\eta)\sin\varphi,\sqrt{\eta(2-\eta)}r\Big).
\end{equation}
From here, we calculate the components of the metric tensor and zweibein. Then, the Hamiltonian on the left side of (\ref{DirEq}) will be \cite{sitenko}
\begin{equation}\hat{H}_s=\hbar v\left\{{\rm i}\sigma^2\partial_r-\sigma^1r^{-1}\left[(1-\eta)^{-1}\left({\rm i}s\partial_{\varphi}+\frac{3}{2}\eta\right)+
\frac{1}{2}\sigma^3\right]\right\},\hspace{1cm}s=\pm 1,\end{equation}
the solution is given by the Bessel functions. Here, $v$ represents the Fermi velocity which gives the velocity of the electrons corresponding to the Fermi energy \cite{pincak}.\\

\subsection{Influence of the spin--orbit interaction}\

\begin{figure}[h]
\centering
\includegraphics[width=110mm]{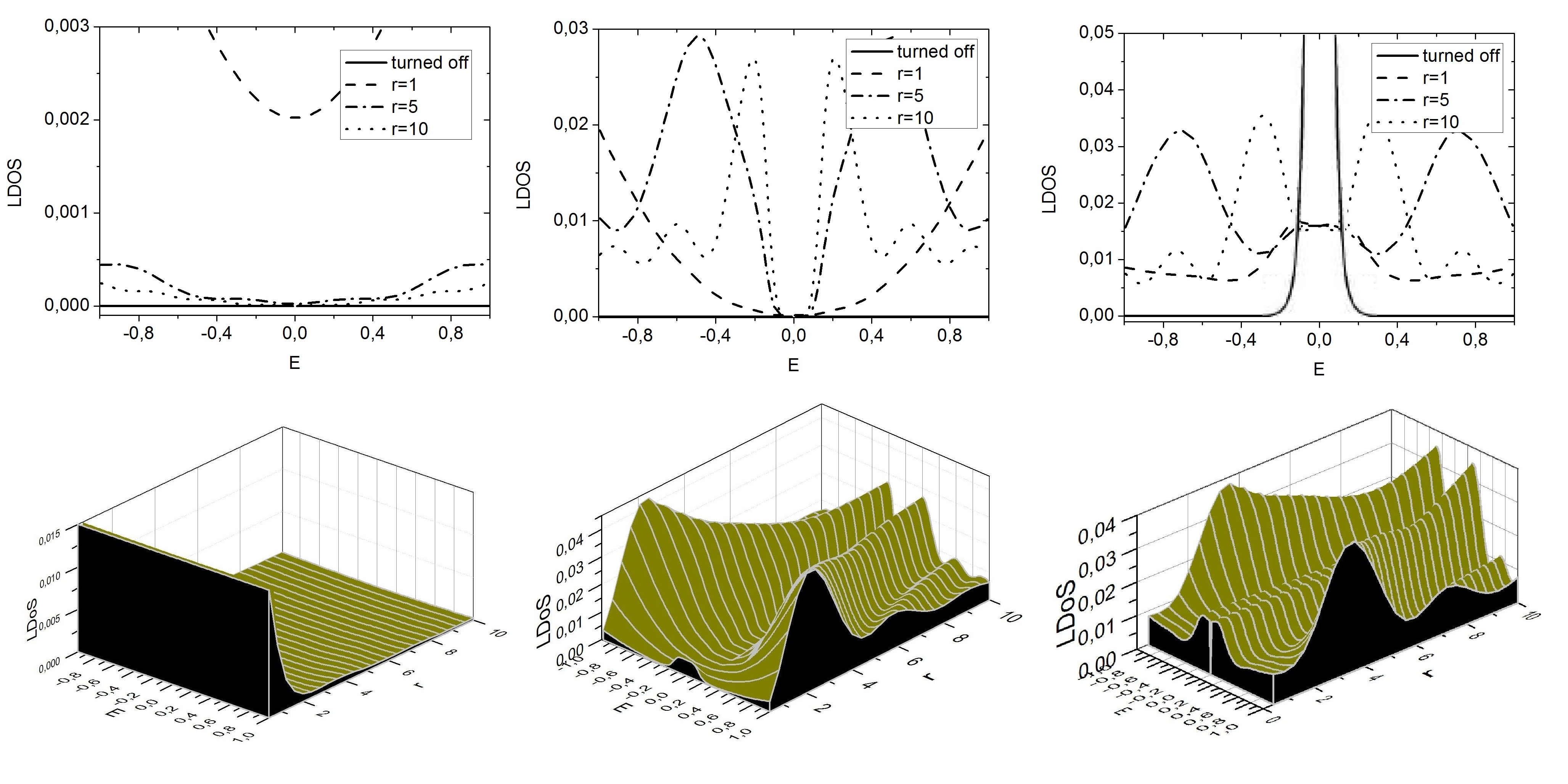}
\caption{The LDoS of the graphitic nanocone influenced by the spin--orbital interaction for $n=0$ and different distance from the tip in the 2D and 3D graphs: 1 defect in the tip (left), 2 defects (middle), 3 defects (right). ``Turned off'' denotes the case when the influence of the SOI is excluded.}\label{LDOSspin}
\end{figure}

In contrast to the case of the plain graphene, negligible influence of the spin--orbit interaction (SOI) can be considered in the graphitic nanocone. To investigate this influence, we perform  the following substitutions in the Hamiltonian:
\begin{equation}\partial_r\rightarrow\partial_r\otimes I-{\rm i}\frac{\delta\gamma'}{4\gamma R}\otimes\sigma_x(\vec{r}),\hspace{5mm}
{\rm i}\partial_{\varphi}\rightarrow{\rm i}\partial_{\varphi}\otimes I+s(1-\eta)A_y\otimes\sigma_y,\end{equation}
where $\delta$ is the parameter which characterizes the strength of SOI, $\gamma, \gamma'$ are the nearest-neighbor and the next-nearest-neighbor hopping integrals, respectively, $A_y$ is the intrinsic term of the SOI. The other parameters are described in \cite{soc}. We are looking for the solution in the factorized form which is common for all the structures with the radial symmetry:
\begin{equation}\psi(r,\varphi)=e^{{\rm i}n\varphi}\left(\begin{array}{c}f_{n\uparrow}(r)\\
f_{n\downarrow}(r)\\g_{n\uparrow}(r)\\g_{n\downarrow}(r)\end{array}\right).\end{equation}
Here, the radial part must satisfy
\begin{equation}\left(\begin{array}{cccc}0 & 0 & \partial_r+\frac{F}{r} & -\frac{\rm i}{r}C\\0 & 0 & -\frac{\rm i}{r}D &
\partial_r+\frac{F}{r}\\ -\partial_r+\frac{F-1}{r} & \frac{\rm i}{r}D & 0 & 0\\
\frac{\rm i}{r}C & -\partial_r+\frac{F-1}{r} & 0 & 0\end{array}\right)\left(\begin{array}{c}f_{n\uparrow}(r)\\
f_{n\downarrow}(r)\\g_{n\uparrow}(r)\\g_{n\downarrow}(r)\end{array}\right)=E\left(\begin{array}{c}f_{n\uparrow}(r)\\
f_{n\downarrow}(r)\\g_{n\uparrow}(r)\\g_{n\downarrow}(r)\end{array}\right)\end{equation}
with $F=\frac{sn}{1-\eta}-\frac{3}{2}\frac{\eta}{1-\eta}+\frac{1}{2}, C=\xi_x-\xi_y, D=\xi_x+\xi_y$,
$\xi_x$ and $\xi_y$ are the spin--orbital coefficients. For finding the solution, we use the numerical methods. As the result of our calculations, the $LDoS$ is outlined in Fig. \ref{LDOSspin} for $n=0$.\\

\subsection{Green function approach}\

The numerical solution of the LDoS found in the last section is distorted by the uncertainties coming from the systematical errors of the used method of performing the solution. To get more precise results, we can try to find the solution using the Green function approach: in this approach, the LDoS can be evaluated as
\begin{equation}\label{TrGr}
\mathrm{LDoS}(E)=\frac{1}{\pi}\,{\rm Im}\,{\rm Tr}\,G(E-{\rm i}0),\hspace{5mm}G=G_0+\int G_0\hat{V}G_0+\int G_0\hat{V}G_0\hat{V}G_0+\cdots,
\end{equation}
where $G$ denotes the case with SOI, $\hat{V}$ stands for the interaction potential of SOI, and $G_0$ representing the case without SOI has the form
\begin{equation}G_{0}(r'',\varphi'',r',\varphi';E)=\frac{1}{2\pi}\sum\limits_{n\in\mathcal{Z}}e^{i(n+s/2)(\varphi'' -\varphi')}1\otimes\left(\begin{array}{cc}a_{n}^{11}(r'',r') & a_{n}^{12}(r'',r')\\
a_{n}^{21}(r'',r') & a_{n}^{22}(r'',r')\end{array}\right);\end{equation}
$a_{n}^{kl}(r'',r'),\,\,k,l=1,2$ are computed in \cite{sitenko}.

Then, the first order approximation of the Green function has the form
\[\left(G_1\right)_{k,l}=
\frac{\hbar\upsilon\delta}{8\pi}\frac{\gamma'}{4\gamma}\frac{\sqrt{\eta(2-\eta)}}{(1-\eta)^2}\left(\begin{array}{cc}-1 & 0\\ 0& 1\end{array}\right)\sum\limits_{n\in\mathcal{Z}}
\int\limits_0^{+\infty}(a_n^{l1}(r,r')a_{n+1}^{2k}(r',r)-
a_n^{l2}(r,r')a_{n+1}^{1k}(r',r))\,{\rm d}r+
\]
\begin{equation}\label{EqG1}+\,{\rm i}\cdot\frac{\hbar\upsilon\delta ps}{4\pi}\frac{\sqrt{\eta(2-\eta)}}{(1-\eta)^2}\left(\begin{array}{cc}0 & 1\\-1 & 0\end{array}\right)\cdot
\sum\limits_{n\in\mathcal{Z}}\int\limits_0^{\infty}\left(a_n^{l2}(r,r')a_{n}^{1k}(r',r)+
a_n^{l1}(r,r')a_{n}^{2k}(r',r)\right)\,{\rm d}r.\end{equation}
The trace of this matrix is zero for arbitrary $k, l$ and, consequently, from (\ref{TrGr}) follows the zero value of the corresponding contribution to the LDoS. That is why we need to calculate at least the second order approximation. For this purpose, we derive a recursion equation
\begin{equation}(G_{m+1})_{i,j}=\int\left((G_m)_{i,1}\cdot A_{1j}+(G_m)_{i,2}\cdot A_{2j}\right),\end{equation}
where $A_{ij}$ are the elements of the matrix
\begin{equation}\hat{V}G_0=\left(\begin{array}{cc}A_{11} & A_{12}\\ A_{21} & A_{22}\end{array}\right).\end{equation}
As an example, we introduce the corresponding expression for $k=1, l=1, m=1$:
{\footnotesize%\scriptsize
	\[(G_{m+1})_{1,1}=\frac{\hbar v}{2\pi}\delta\sqrt{\eta(2-\eta)}\sum\limits_{n\in\mathcal{Z}}\int\limits_0^{2\pi}
\int\limits_0^{\infty}e^{{\rm i}(n+\frac{s}{2})(\varphi-\varphi')}
\left[-a_n^{12}(r,r')(G_{m,s})_{1,1}\left(\begin{array}{cc}-{\rm i}\frac{\gamma'}{4\gamma}\sin\varphi &
{\rm i}\frac{\gamma'}{4\gamma}\cos\varphi-{\rm i\cdot 2ps}\\{\rm i}\frac{\gamma'}{4\gamma}\cos\varphi+{\rm i\cdot 2ps} & {\rm i}\frac{\gamma'}{4\gamma}\sin\varphi\end{array}\right)\right]\,{\rm d}r\,{\rm d}\varphi+\]
 \begin{equation}+\frac{\hbar v}{2\pi}\delta\sqrt{\eta(2-\eta)}\sum\limits_{n\in\mathcal{Z}}\int\limits_0^{2\pi}
\int\limits_0^{\infty}e^{{\rm i}(n+\frac{s}{2})(\varphi-\varphi')}
\left[a_n^{11}(r,r')(G_{m,s})_{1,2}\left(\begin{array}{cc}-{\rm i}\frac{\gamma'}{4\gamma}\sin\varphi &
{\rm i}\frac{\gamma'}{4\gamma}\cos\varphi+{\rm i\cdot 2ps}\\{\rm i}\frac{\gamma'}{4\gamma}\cos\varphi-{\rm i\cdot 2ps} & {\rm i}\frac{\gamma'}{4\gamma}\sin\varphi\end{array}\right)\right]\,{\rm d}r\,{\rm d}\varphi.\end{equation}}
Then, using the usual procedure we get for the trace of the matrix $G_2$
{\small
\begin{equation}\label{EqG2}{\rm Tr}\,G_2=-\frac{\hbar^2v^2}{16\pi^3}\frac{\delta^2\eta(2-\eta)}{(1-\eta)^3}\left[\left(\frac{\gamma'}{4\gamma}\right)^2
\sum\limits_{|n_{01}-n_{12}|=1}\int\limits_0^{\infty}
\int\limits_0^{\infty}I_1(r',r_1,r_2)\,{\rm d}r_1\,{\rm d}r_2-8p^2\sum\limits_{n\in\mathcal{Z}}\int\limits_0^{\infty}
\int\limits_0^{\infty}I_2(r',r_1,r_2)\,{\rm d}r_1\,{\rm d}r_2\right],
\end{equation}}
where
\[I_1(r',r_1,r_2)=a_{n_{01}}^{12}(r_2,r')\left(a_{n_{12}}^{12}(r_1,r_2)a_{n_{01}}^{11}(r',r_1)-
a_{n_{12}}^{11}(r_1,r_2)a_{n_{01}}^{21}(r',r_1)\right)+\]
\[+a_{n_{01}}^{11}(r_2,r')\left(a_{n_{12}}^{21}(r_1,r_2)a_{n_{01}}^{21}(r',r_1)-
a_{n_{12}}^{22}(r_1,r_2)a_{n_{01}}^{11}(r',r_1)\right)+\]
\[+a_{n_{01}}^{22}(r_2,r')\left(a_{n_{12}}^{12}(r_1,r_2)a_{n_{01}}^{12}(r',r_1)-
a_{n_{12}}^{11}(r_1,r_2)a_{n_{01}}^{22}(r',r_1)\right)+\]
\begin{equation}+a_{n_{01}}^{21}(r_2,r')\left(a_{n_{12}}^{21}(r_1,r_2)a_{n_{01}}^{22}(r',r_1)-
a_{n_{12}}^{22}(r_1,r_2)a_{n_{01}}^{12}(r',r_1)\right),\end{equation}
\[I_2(r',r_1,r_2)=a_{n}^{12}(r_2,r')\left(a_{n}^{12}(r_1,r_2)a_{n}^{11}(r',r_1)+
a_{n}^{11}(r_1,r_2)a_{n}^{21}(r',r_1)\right)+\]
\[+a_{n}^{11}(r_2,r')\left(a_{n}^{22}(r_1,r_2)a_{n}^{11}(r',r_1)+
a_{n}^{21}(r_1,r_2)a_{n}^{21}(r',r_1)\right)+\]
\[+a_{n}^{22}(r_2,r')\left(a_{n}^{12}(r_1,r_2)a_{n}^{12}(r',r_1)+
a_{n}^{11}(r_1,r_2)a_{n}^{22}(r',r_1)\right)+\]
\begin{equation}+a_{n}^{21}(r_2,r')\left(a_{n}^{22}(r_1,r_2)a_{n}^{12}(r',r_1)+
a_{n}^{21}(r_1,r_2)a_{n}^{22}(r',r_1)\right).\end{equation}

Now we can use (\ref{TrGr}) to calculate the LDoS  with the precision up to the second order of the Green function approach:
\begin{equation}\mathrm{LDoS}_2(E)=\frac{1}{\pi}\,{\rm Im}\,{\rm Tr}\,\left[G_0(E-{\rm i}0)+G_1(E-{\rm i}0)+G_2(E-{\rm i}0)\right],\end{equation}
where the lower index $2$ in LDoS$_2$ means the mentioned precision. For the purpose of comparison with our results from \cite{soc}, we use the values from the third graph in Fig. \ref{LDOSspin}, i. e. 3 defects, $n=0$ and $r'=5$ (in the mentioned figure, $r$ is used instead of $r'$).

To calculate $G_0$, we exploit the expression (A.6) in \cite{sitenko} for the calculation of ${\rm Tr}\,G_0$ :
\begin{equation}\frac{E}{\pi}\int_0^{\infty}\frac{p\,{\rm d}p}{\hbar^2v^2p^2-E^2}\sum\limits_{n\in\mathcal{Z}}
[J_{2sn-1}(pr)J_{2sn-1}(pr')+J_{2sn}(pr)J_{2sn}(pr')],\end{equation}
where $r\rightarrow r'$. In the sum, we use only the term $J_{2sn}(pr)J_{2sn}(pr')$ for $n=0$. Tr\,$G_1$ is zero for arbitrary $n$, as follows from (\ref{EqG1}). Tr\,$G_2$ is calculated using (\ref{EqG2}).\\

\begin{figure}[h]
\centering
\includegraphics[width=45mm]{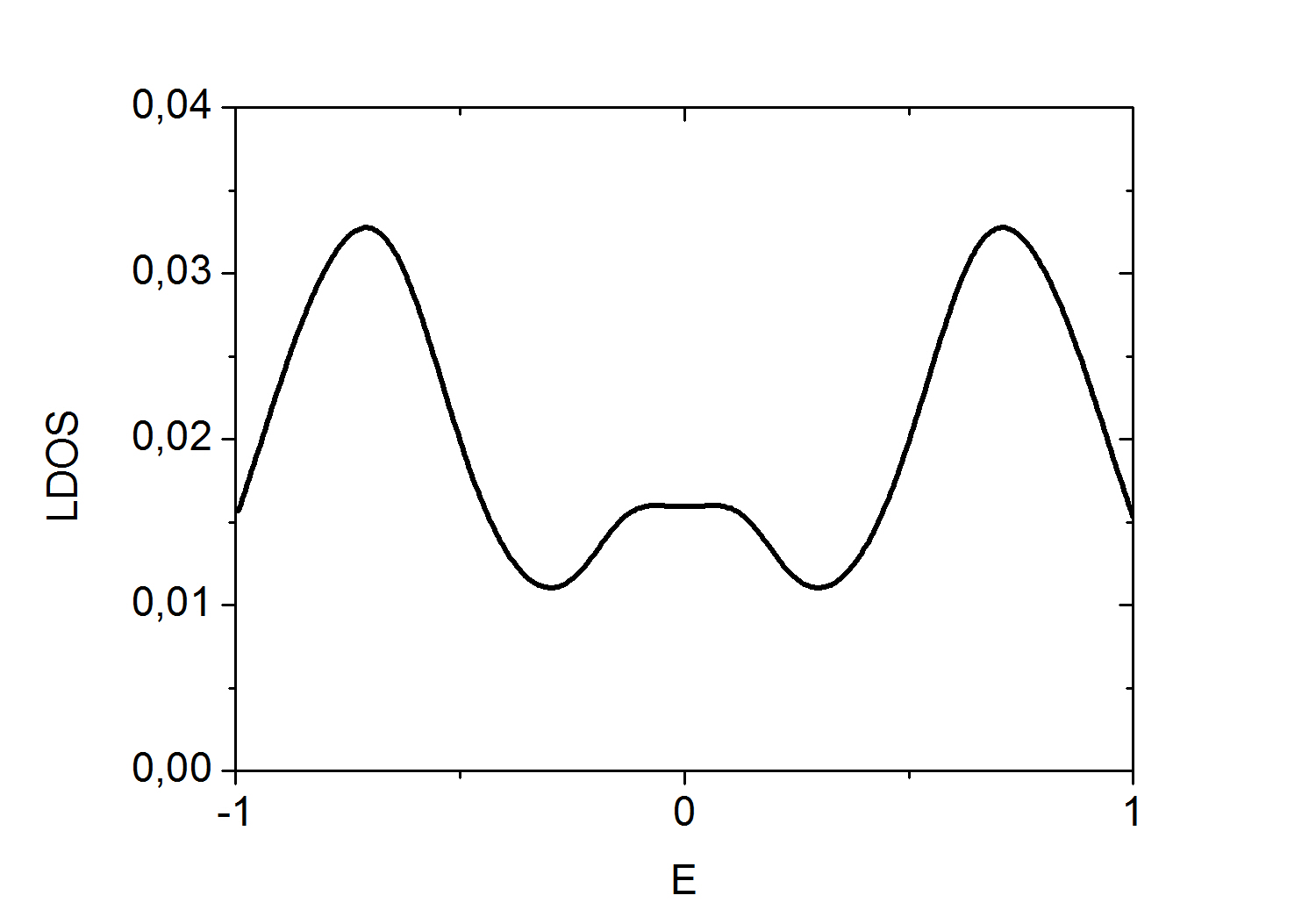}
\includegraphics[width=45mm]{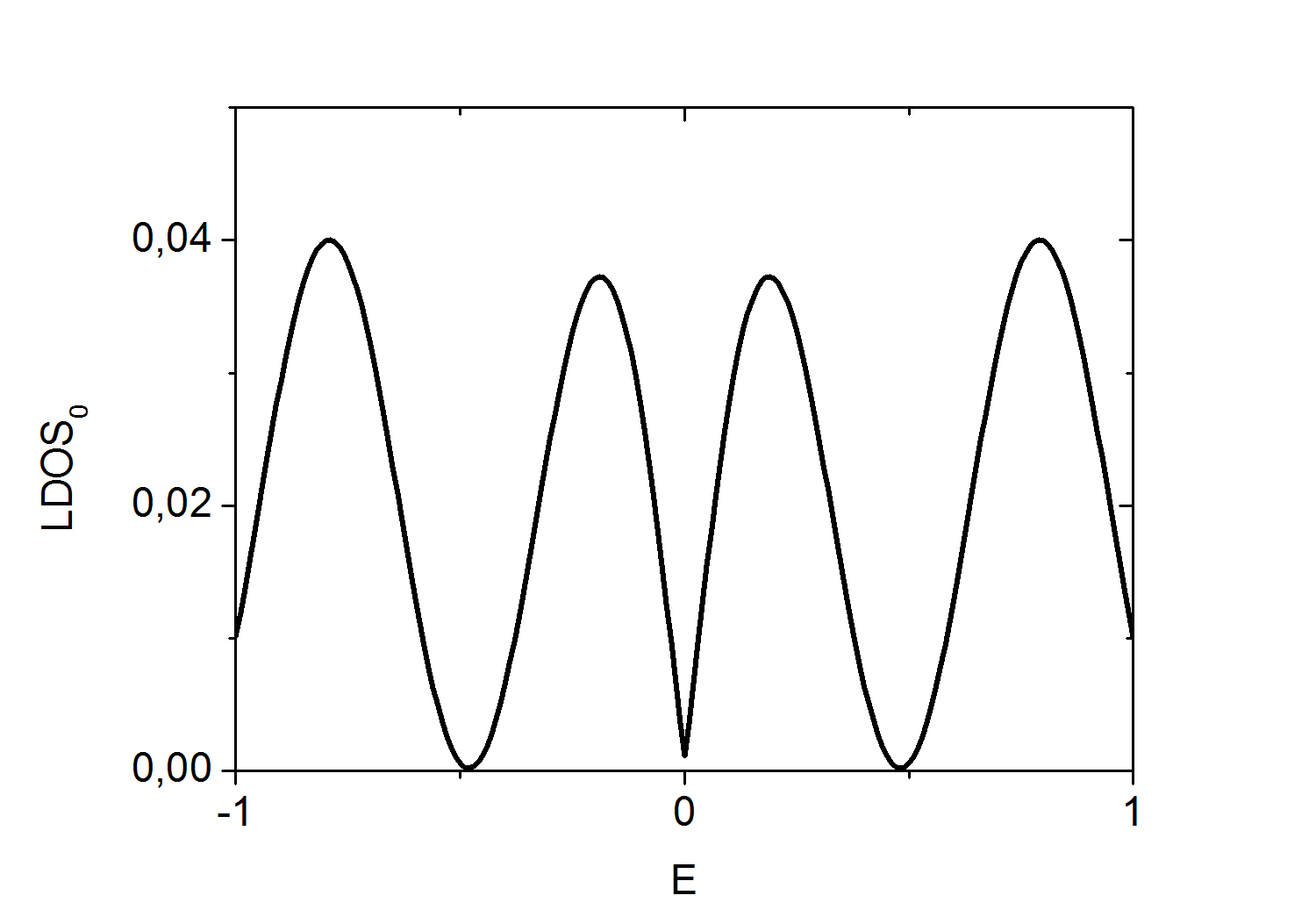}
\includegraphics[width=45mm]{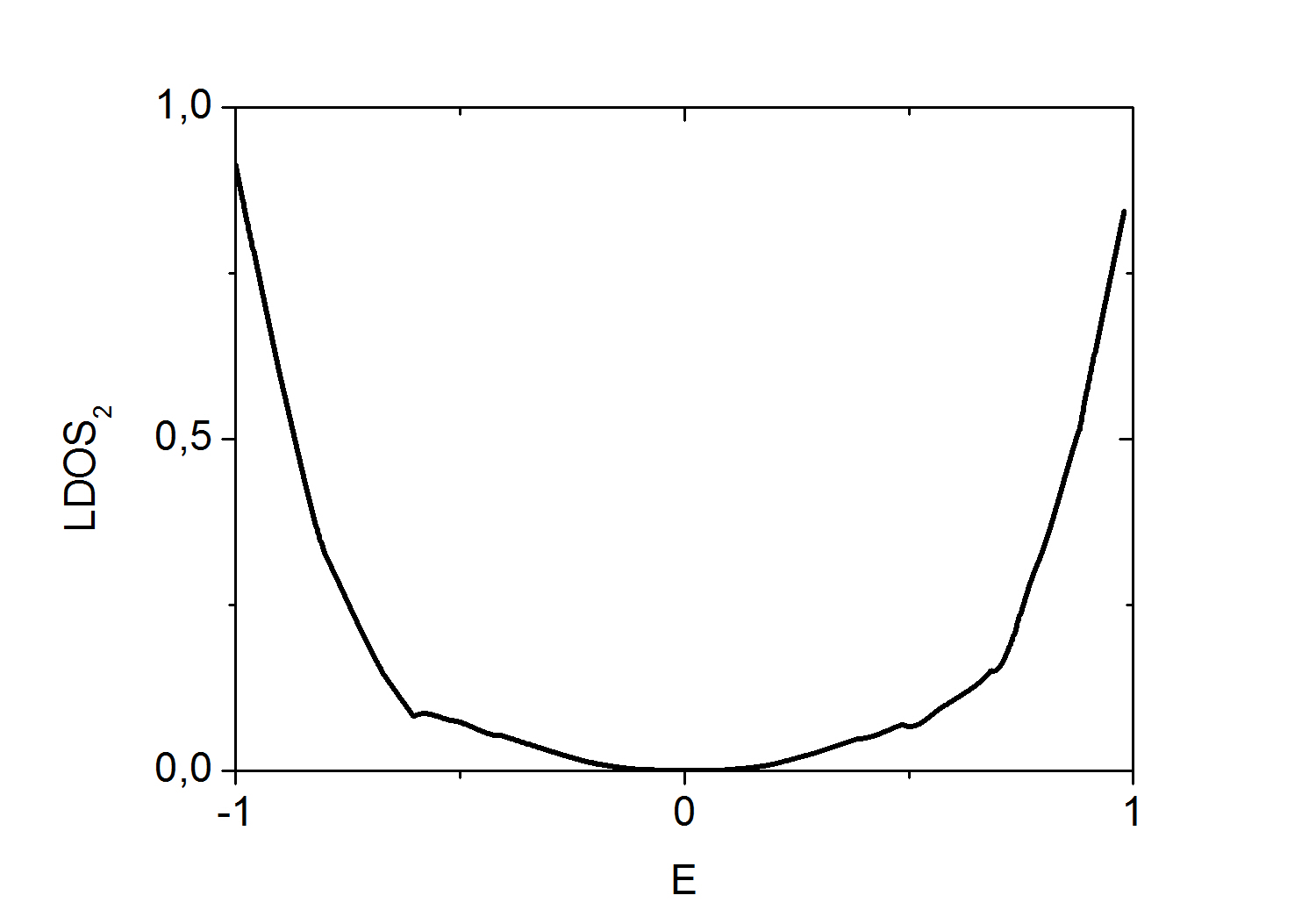}
\caption{LDoS calculated using a numerical method in \cite{soc} (left), the zeroth order approximation calculated using the relations in \cite{sitenko} (middle) and the second order approximation calculated using (\ref{EqG2}) (right), performed for the case of 3 defects and the values $n=0$, $r'=5$}\label{LDOS_Green}
\end{figure}

\section{Conclusion and discussion}\

In Fig. \ref{LDOS_Green}, we see the graph of the LDoS from the right part of Fig.~\ref{LDOSspin} for the mentioned values and the corresponding 0-th order and the second order approximation calculated using the Green function. If the second order approximation of the Green function is sufficient, the sum of the middle part and of the right part of Fig.~\ref{LDOS_Green} should give the left part of this figure. But as we can see, the scale of the zeroth order approximation is comparable with the scale of the supposed result, while the scale of the second order approximation differs significantly. This is caused at least by one of these 2 factors: either the second order approximation is not sufficient to get the correct result, or the results for the second order are not normalized sufficiently. From a simple comparison of the graphs one could expect the first possibility, but only the calculation of the next orders and the performance of the corresponding normalization can confirm this estimation. The work in this direction is still in progress.

\vskip 0.2cm ACKNOWLEDGEMENTS --- This paper was supported by the Science and Technology Assistance Agency under
Contract No. APVV-0171-10, VEGA Grant No. 2/0037/13 and Ministry of Education Agency for Structural Funds
of EU within the project 26220120021, 26220120033 and 26110230061.


\begin{thebibliography}{99}

\bibitem{wallace}
P. R. Wallace, Phys. Rev. \textbf{71}, 622 (1947)

\bibitem{slonczewski}
J. C. Slonczewski and P. R. Weiss, Phys. Rev. \textbf{109}, 272 (1958)

\bibitem{mele}
D. P. DiVincenzo and E. J. Mele, Phys. Rev. B \textbf{29}, 1685 (1984)

\bibitem{kochetov}
E. A. Kochetov, V. A. Osipov and R. Pincak, J. Phys.: Condens. Matter \textbf{22}, 395502 (2010)

\bibitem{smotlacha}
J. Smotlacha, R. Pincak and M. Pudlak, Eur. Phys. J. B \textbf{84}, 255 (2011)

\bibitem{fecko}
M. Fecko, \textit{Differential Geometry and Lie Groups for Physicists}, (Cambridge University Press, New York, 2006), 417 pp.  %79

\bibitem{sitenko}
Yu. A. Sitenko and N. D. Vlasii, Nucl. Phys. B \textbf{787}, 241 (2007)

\bibitem{pincak}
R. Pincak and M. Pudlak, chapter in \textit{Progress in Fullerene Research}, (Nova Science Publishers, New York, 2007)

\bibitem{soc}
R. Pincak, J. Smotlacha and M. Pudlak, Eur. Phys. J. B \textbf{88}, 17 (2015)

\end{thebibliography}
\end{document}